# Short-Range Order Structure Motifs Learned from an Atomistic Model of a $Zr_{50}Cu_{45}Al_5$ Metallic Glass

Jason J. Maldonis[1], Arash Dehghan Banadaki[2], Srikanth Patala[2], Paul M. Voyles[1]
[1]Department of Materials Science and Engineering, University of Wisconsin-Madison, Madison, WI, USA
[2]Department of Materials Science and Engineering, North Carolina State University, Raleigh, NC, USA

Keywords: Simulation, Machine Learning, Metallic Glass, Structure, Short-Range Order, Motif

**Abstract**

The structural motifs of a $Zr_{50}Cu_{45}Al_5$ metallic glass were learned from atomistic models using a new structure analysis method called *motif extraction* that employs point-pattern matching and machine learning clustering techniques. The motifs are the nearest-neighbor building blocks of the glass and reveal a well-defined hierarchy of structures as a function of coordination number. Some of the motifs are icosahedral or quasi-icosahedral in structure, while others take on the structure of the most close-packed geometries for each coordination number. These results set the stage for developing clearer structure-property connections in metallic glasses. Motif extraction can be applied to any disordered material to identify its structural motifs without the need for human input.

**Introduction**

Local structure in metallic glass (MG) is characterized by large coordination numbers, efficient packing [1], and a rich variety of structures [2–4]. The large coordination numbers result in complex short-range order (SRO) structures, and the packing of these structures is frustrated by their variety and shapes. This variety and frustration leads to significant structural disorder [5,6] that impedes our ability to identify useful abstractions of MG structure. SRO structure in MGs was initially modeled as the dense packing of monotonic hard-spheres [7]; however, this model was unable to explain experimental data that supported the existence of chemical order [8,9] and was replaced by theories incorporating information about the different atomic species present in the glass.

The efficient cluster packing (ECP) model [1,10,11], for example, incorporated chemical information by applying packing constraints using specific atomic radii ratios and assuming negative heats of mixing between the atomic species [12]. Theories such as ECP that assume efficient packing of atoms provide constraints on the local structure of MGs that allow researchers to derive optimally packed structures at the SRO length scale. For example, using polytetrahedral packing theories, Frank and Kasper [13,14] and Y.Q. Cheng and E. Ma [15] identified efficiently packed polytetrahedral clusters ("Z-clusters") with coordination numbers (CNs) ranging from 8 to 16. Other idealized structures were identified using energy minimization [16,17] or alignment techniques [18], and close-packed medium-range order (MRO) structures such as Bergman [19] and Mackay [20] polyhedra have also been suggested [21–24]. Many of these idealized structures are found in various simulated structures of MGs [3,15,25–27], and in some instances these structures correlate with properties such as glass-forming ability [3,27–29], dynamic heterogeneities [30,31], atomic mobility [27,32], or deformation behavior [33,34].

Despite these successes, scientists still cannot reliably design MGs with specific properties using only knowledge of the alloying components and processing conditions. Understanding and predicting the glass-forming ability of a new alloy is a complex, multifaceted problem, and part of that complexity arises from inadequate understanding of the structure of glasses and liquids. In particular, the short-range disorder within the unit defined by an atom and its nearest neighbors (henceforth called a "SRO unit" rather than a "cluster" to avoid confusion with machine learning clustering terminology used later on) manifests as displacements around preferred atomic sites due to thermal vibrations or longer-range packing constraints. At the MRO length scale, the disorder is created by the rotational degrees of freedom of the SRO units as they pack together, which disrupts long-range order entirely. Together, these forms of disorder make quantifying the SRO structure of MGs difficult because rotation-invariant analysis techniques are prohibited by the disorder in the atom positions, and the rotations of the SRO units makes direct comparison of the atom positions ineffective.

The most common approach to circumventing these problems of disorder and analyzing the structure of MGs from atomistic models is the use of topological analysis techniques such as Voronoi analysis [35] and common neighbor analysis [36]. The Voronoi index analysis abstracts the exact atomic positions of SRO units into shapes, and describes the shapes qualitatively and independent of orientation. In this formalism, a polygon is constructed that represents the volume that "belongs to" an atom, in the sense that every point in the volume is closer to the central atom than to any other atom in the structure. This polygon is then abstracted into a set of indices, $<n_3\ n_4\ n_5\ n_6\ ...>$, designating the number of faces of the polygon with 3, 4, 5, 6, … edges. The indices are used as a characterization of the atom's local environment. This Voronoi index (VI) method enables a binary metric of similarity for the local structure around two atoms (i.e. *do these two atoms have the same VI, yes or no?*).

VIs provide an abstract description of the SRO units in MG structure, and in Zr-Cu-based MGs SRO units with icosahedral VI <0 0 12 0> become more populous and kinetically slower [37] as the material undergoes its glass transition [38,39]. The increasing fraction and kinetic slowdown of icosahedral VI as the temperature approaches $T_g$ provide a structural description for the dynamical arrest that occurs during the glass transition; it is hypothesized that a structural "backbone" of icosahedra forms that increases in size until a percolation threshold is reached, and the slow properties of the icosahedral network arrests and largely immobilizes the remaining structure [37,40]. Icosahedra therefore form the basis for a strong structure-property connection in Zr-Cu-based MGs.

Despite the successes of Voronoi analysis in Zr-Cu-based MGs, the binary yes/no and topological nature of categorization techniques has some drawbacks. VI provide a way to measure whether structures are the same, but no quantitative way to measure whether they are *similar*. As a result, there is a long and growing list of VIs in the literature that are called "quasi-icosahedral", meaning variously that they have a lot of five-edge faces, or that one can imagine moving one or two atoms a small distance to create a perfect icosahedron, or some other, even less well-defined criterion [3,26,37,41–43]. The topological nature is both a benefit, since it confers a level of abstraction that has limited sensitivity to the details of atomic positions and rotations, but also a hindrance, since it is precisely the details of the atom positions that determine the strength of the interatomic interactions.

In Ref. [44] we presented a geometric approach to categorizing glass SRO units (amongst other structures like grain boundaries) that yields a quantitative metric of similarity between two structures. The ability to compute a degree of similarity, or structural distance, between two SRO units solves one of the difficulties with VI analysis, and using atom positions that preserve geometry mitigates the other. In addition, similarity information enables the application of tools based on metric spaces for additional analysis of MG structure. Here we apply density-based clustering [45], a machine learning method based on metric spaces, to learn idealized, important SRO unit structures from atomistic models of a $Zr_{50}Cu_{45}Al_5$ metallic glass.

A geometric similarity metric comparing two SRO units necessitates a solution to the rotational variance of the SRO units in a MG structure. Recently, we adapted a technique called point-pattern matching (PPM) [44] from the computer vision literature to the study of the 3D structure of materials. PPM works by aligning two sets of 3D points into as similar of an orientation and position as possible using an approximate rigid graph registration technique that can handle mild disorder between the two structures. After alignment, the similarity of the structures can be compared using any geometric metric. A similar structure analysis technique was developed by Fang *et al.* [18,24,46]. Their *cluster alignment* technique applies molecular dynamics with two potentials—one to constrain the bond distances within a SRO unit and the other to encourage alignment of atom positions in different SRO units—to collectively align a group of SRO units. The result of the cluster alignment method is a set of SRO units, all in similar orientation. Using this collective alignment, a probability density map describes the most probable atom positions of the aligned SRO units. The PPM approach differs by enabling fast alignment of any two structures, rather than a global alignment of many structures. The PPM method facilitates quantification of similarity between all pairs of SRO units in a material, independent of a collective alignment.

In this work, we used PPM to quantify the similarity between all pairs of SRO units in a $Zr_{50}Cu_{45}Al_5$ MG. These dissimilarity scores function as the equivalent of a distance metric between two points in multidimensional structural configuration space, which is the data needed for machine learning clustering algorithms to cluster similar sets of SRO units. In this way, we learn all of the important classes of SRO structures in the MG directly from the atom position data, without human intervention. In $Zr_{50}Cu_{45}Al_5$, the algorithm identified thirty motifs, including an icosahedral motif and structures similar to other hypothesized SRO structures with different coordination numbers. After identification of the motifs, we discuss their structure, chemistry, and stability as a function of temperature.

**Methods**

*Molecular Dynamics*

A $Zr_{50}Cu_{45}Al_5$ liquid with 9826 atoms in a cubic box with periodic boundary conditions was equilibrated at 2000 K using molecular dynamics (MD) [47,48] with an EAM potential [49] in an NPT ensemble, then cooled to 50 K at $5\times10^{10}$ K/s. The simulation timestep was 0.1 fs, a Nose-Hoover thermostat and barostat were used, and the initial model configuration was created by equilibrating a bcc lattice with the correct composition in an NVE ensemble for 20 ps. The glass's $T_g$ was ~ 740 K. Snapshots of the trajectory were extracted every 50 K, from 1850 K to 50 K, and their inherent structures were derived using conjugate gradient minimization [50].

The CN distribution and partial pair distribution functions, $g(r)$, are shown in Figure 1a and b for the 600 K model. These results are consistent with other simulation [49] and experimental [51] work on similar compositions. Note the high variety of center atoms types in SRO units with CN 12 in Figure 1a. VI statistics are shown in Figure 1c for the liquid (1600 K) and glass (600 K) and are also similar to previous results [49].

*Creation of the Z-Cluster Structures*

For comparison to the learned metallic glass motifs, we created models of the generalized Frank-Kasper polyhedra called "Z-clusters" [15]. The Z-clusters used in this work are efficiently packed polytetrahedral structures given the radii constraints in Table 1 in Ref [52]. They were created using a combination of MD and conjugate gradient simulations and the methods used in Refs [15] and [52]. The radii in Table 1 in Ref [52] were used to create a binary Lennard-Jones (LJ) potential for each CN. The two atomic species were given a stronger interaction potential to keep the neighboring atoms (species A) bonded to the center atom (species B). The approximate structure of the Z-clusters that share a CN with the Frank-Kasper polyhedra were generated by hand, then the bond lengths were optimized using conjugate gradient minimization with the appropriate LJ potential for that CN in LAMMPS. We use "Z$n$" to denote the Z-cluster with CN $n$, so Z9 is the Z-cluster with CN 9, *etc*.

The Z11 and Z13 structures proved especially difficult to create, likely because there is not an obvious close-packed polytetrahedral structure for these CNs. In fact, the most close-packed structure for these CNs may not be polytetrahedral. As a result, Z11 and Z13 are not perfectly polytetrahedral, although we believe them to be optimally close-packed given the radii constraints. Z13 is nearly polytetrahedral, except that the tetrahedra are slightly distorted. Z11 has one atom that disrupts the polytetrahedral structure but preserves the symmetry of the structure. Z11 and Z13 were created by quenching a LJ glass using the appropriate LJ potential, and the motif extraction technique (see below) was applied to SRO units with CN 11 and 13 and with VI <0 2 8 1> and <0 1 10 2>, respectively. The minimal disorder in the resulting motifs was further minimized using conjugate gradient minimization.

*Motif Extraction*

The motif extraction method is illustrated in Figure 2. The nearest neighbors of each atom in an atomic model were identified using a radial cutoff of 3.6 Å, the minimum between the first and second peaks in the total pair distribution function. This cutoff value does not change with temperature and was used for all trajectory snapshots. Each atom and its nearest neighbors define a SRO unit. Each SRO unit was radially contracted until the average bond length between the center atom and all neighboring atoms was equal to 1.0. Initially, PPM alignment was performed on all possible pairs of SRO units. However, the primary result of that alignment was to sort the units by CN, so for the analysis presented here, the SRO units were first separated into groups corresponding to their CN, then PPM was performed on each pair of SRO units with the same CN. The PPM alignment process of two SRO units required ~ 200 ms on modern hardware, but all of the alignments are independent, making the total alignment process embarrassingly parallel. An implementation of PPM that takes advantage of this parallelization is available on GitHub at https://github.com/spatala/ppm3d.

After alignment, four metrics were used to quantify the dissimilarity of the two structures:

$$L^2 = \frac{1}{n}\sqrt{\sum_{i=0}^{n}(A_{i_x} - B_{i_x})^2 + (A_{i_y} - B_{i_y})^2 + (A_{i_z} - B_{i_z})^2}$$

where $A_{i_x}$ is the x-coordinate of atom $i$ in the SRO unit $A$, etc., and $n$ is lower of the CN of $A$ or $B$;

$$L^1 = \frac{1}{n}\sum_{i=0}^{n}|A_{i_x} - B_{i_x}| + |A_{i_y} - B_{i_y}| + |A_{i_z} - B_{i_z}|;$$

$$L^\infty = \max\left(|A_{i_x} - B_{i_x}| + |A_{i_y} - B_{i_y}| + |A_{i_z} - B_{i_z}|\right);$$

and

$$L^{\measuredangle} = \frac{1}{m}\sum_{i,j}|\measuredangle(A_i, A_j) - \measuredangle(B_i, B_j)|$$

where the function $\measuredangle(A_i, A_j)$ calculates the angle between atom $A_i$ and atom $A_j$ through the center of the SRO unit, $m$ is the total number of bonds in $A$, $i$ and $j$ are indices of neighboring atoms in $A$, $B_i$ is the atom in $B$ that corresponds to atom $A_i$ in $A$ after alignment, and the summation runs over the indices of all pairs of neighbors in $A$. $L^{\measuredangle}$ is therefore a measure of the mean angular dissimilarity of $A$ and $B$. The distribution of values of each of these metrics were normalized to have a mean and standard deviation of 1.0, then the geometric mean of these four normalized metrics,

$$D = \sqrt[4]{L^2_{norm} \cdot L^1_{norm} \cdot L^\infty_{norm} \cdot L^{\measuredangle}_{norm}},$$

was computed. From the dissimilarity, one could also calculate a similarity metric as defined in [53] from $S = e^{-D}$.

$D$ values from the pairs of SRO units with the same CN were used to form separate dissimilarity matrices, one for each CN. HDBSCAN [54,55], a machine learning clustering algorithm, was applied recursively to each dissimilarity matrix until the resulting clusters were primarily classified as one noisy cluster. Two properties of HDBSCAN make it especially well-suited to cluster SRO units: first, it is a spatial clustering algorithm that clusters data based on the local density of points and can therefore identify clusters with non-spherical shapes in $d$-dimensional configuration space. This is advantageous because we have a poor understanding of the shape of the data in $d$-dimensional space. Second, HDBSCAN has a well-defined notion of noise and can classify points as outliers. Therefore, SRO units with exceptionally strong disorder will be classified as outliers, and their unusual local environments do not influence the identified clusters. (In the context of MG structure, these outlying SRO units may constitute an interesting way to define the concept of a defect.) In addition, the hierarchical nature of HDBSCAN allows for a recursive implementation that can identify clusters with different local densities while automatically identifying the optimal number of clusters.

The result of the recursive HDBSCAN algorithm is one or more clusters of SRO units with the same CN and similar geometry, as defined by the metric $D$. Each cluster of SRO units produced in this way was used to create a motif. First, all SRO units in the cluster were aligned into the same orientation by using PPM to align each SRO unit to the one SRO unit that was most geometrically representative of the group (defined as the SRO unit with the lowest mean dissimilarity score calculated over all SRO units in the same cluster). Once the SRO units were in the same orientation, the atom-to-atom mappings provided by PPM [44] determine which atom in each SRO unit corresponds to each of the $n$ atomic sites, where $n$ is the CN of the SRO units. These "bunches" of atoms around each atomic site represent the disorder of the structure. By averaging the atomic positions in each "bunch," the disorder is averaged out and the underlying structure of a group of similar SRO units is identified.

We call the structure that is produced by averaging the atom positions around each atomic site a *motif*. Each motif is representative of a subset of the SRO units in the original model. One motif is produced per cluster identified via the recursive HDBSCAN algorithm, and collectively these motifs form the basis for the SRO structure of the material. We emphasize that these motifs are learned from the model with no input or prior knowledge from the experimenter. In addition, it is worth noting that while this discussion has focused on SRO units in particular, in principle the motif extraction method can be applied to any sized structures. The motif extraction code is available on GitHub at https://github.com/paul-voyles/motifextraction.

**Results**

The motif extraction technique was performed on models at 600 K, 900 K, 1200 K, and 1500 K during the MD simulation after conjugate gradient minimization in order to test whether the motifs change with temperature and to generate a superset of motifs capable of describing the structure at all temperatures. A vast majority of SRO units had CNs ranging from 8-15 in this temperature range. Many motifs at different temperatures were similar—defined via PPM where "similar" motifs had a $D$ value below a certain (temperature dependent) threshold—, so a subset of the total set of motifs at these four temperatures was chosen to represent the SRO structure of the material at all temperatures. Overall, we identified 30 unique motifs in $Zr_{50}Cu_{45}Al_5$, shown in Figure 3 with orientations chosen to illustrate various symmetry elements in the motifs. We use the notation $n_A$ to label each motif where $n$ is the coordination number of the motif and $\{A, B, C, . . .\}$ enumerates the motifs at constant $n$. Motifs with structure most similar to the Z-cluster with the same CN are labeled with an additional superscripted $Z$. The atomic coordinates of the motifs are provided in the Supplementary Information (SI).

Every SRO unit in each 50 K snapshot in the MD trajectory was aligned to each motif in Figure 3 using PPM, and $D$ was computed for every motif + SRO unit pair. Each SRO unit was assigned to the motif with the lowest $D$ with the same CN as the SRO unit. The number of SRO units assigned to a given motif is henceforth referred to as the *population of the motif* in the atomic model.

*Structure-Property Connections*

Figure 4a shows normalized data of the motif population as a function of temperature. In order to highlight the informative changes in these populations, we normalized the motif populations in three ways. First, we divided the population of each motif by the number of SRO units with the same CN as the motif at each temperature. This decouples the change in motif population from the overall change in CN of the material as it is cooled (Figure 4b). Second, we multiplied by the number of the motifs with the same CN (which varies with CN), which allows for direct comparisons of "populations" of motifs across CNs. That is,

$$P_{rel}(m,T) = \frac{P(m,T)}{P(n=n_m,T)} * M(n=n_m)$$

where $P(m,T)$ is number of SRO units assigned to motif $m$ at temperature $T$, $P(n,T)$ is the population of clusters with CN $n$ at temperature $T$, $M(n)$ is the number of motifs with CN $n$, and $n_m$ is the CN of motif $m$. Finally, the curves were vertically offset so that the mean value of the normalized populations at high temperature was approximately zero:

$$P_{rel-0}(m,T) = P_{rel}(m,T) - \langle P_{rel}(m,T) \rangle_{T \geq 1700}$$

The result of these normalizations allows for a more robust visual comparison of motifs across temperature and CN and highlights the important changes in the populations of the motifs as the material is cooled.

Figure 5 shows the normalized "energy" of each motif. The motifs themselves do not have a well-defined energy, in part because they exist in isolation rather than in the context of a MG environment. To compute the motifs' energies, first the energy of every atom in the inherent structure of the model from which the motif was learned was calculated from the EAM potential used in the MD simulations, then the mean of the energies for the atoms whose SRO unit was assigned to each motif was calculated. The mean energy of the SRO units changes with CN, so in order to compare the relative energies of motifs across CNs, we subtracted the mean energy of all SRO units with the same CN as the motif from the mean energy of all SRO units assigned to a given motif. This difference in energy is plotted in Figure 5, which captures the significantly lower relative energy of motif $12_A{}^Z$ in comparison to all other motifs.

We quantified the chemical order of the 600 K inherent structure by observing the species of atoms at the center and in the shells of the SRO units. Figure 6 shows the relative concentration of (a) the center atom specie and (b) the average shell composition with respect to the composition of the model ($Zr_{50}Cu_{45}Al_5$) as a function of CN. Al atoms have a high tendency to both be at the center and be in the shell of SRO units with CN 12. Figure 7 shows analogous center-atom and shell composition data for each motif. The numbers in Figure 7a are divided by those in Figure 6a, and the numbers in Figure 7b are divided by those in Figure 6b; this allows for a relative comparison of chemistry between motifs with different CNs.

**Discussion**

*Structural Hierarchy of Motifs Similar to Z-clusters in $Zr_{50}Cu_{45}Al_5$*

In their seminal 1958 paper [13], Frank and Kasper described a subset of close-packed, polytetrahedral structures in terms of rings of atoms constrained in a 2D plane. They determined that for structures with CN above 12, 6-atom rings were favorable over 4-atom rings, and 6-atom and 5-atom rings were the identifying characteristics of close-packed, polytetrahedral structures. Many of the motifs identified in this work benefit from this visual description of planar rings in addition to the quantitative PPM metric. We describe the topology of structures with planar rings using notation analogous to "1-5-5-1", which would describe a structure with two planar 5-atom rings (usually rotated with respect to each other) and two single atoms on the "top" and "bottom" of the structure (i.e. an icosahedron). Note that this description is subject to the orientation of the 2D projection, but it nevertheless remains useful.

The thirty motifs' CNs, VIs, and dissimilarities to the Z-cluster with the same CN are shown in Table 1. With the exception of CN 8, each CN has a motif with structure similar to the Z-cluster with the same CN. The motif with CN 8 is dissimilar to Z8 likely because SRO units with CN 8 are unfavorable due to their atomic radii and instead form due to fluctuations of atomic nearest neighbors in the liquid. There are nine motifs for CN 12, more than any other CN, which is likely a result of the large chemical diversity of SRO units with CN 12 (see Figures 1a and 6b).

Figure 8 shows the motif for each CN that is most similar to the Z-cluster with the same CN. The motifs for CN 9-14 follow a clear hierarchy of structure with increasing CN, and the placement of an additional atom in the structure (which increases the CN) is often predictable. The topology of these motifs can be described from lowest to highest CN as 1-4-4, 1-4-4-1, 1-5-4-1, 1-5-5-1, 1-6-5-1, and 1-6-6-1. The single motif with CN 15 ($15_A^Z$) slightly breaks the pattern and has a topology that is 1-6-6-2, possibly because 7-atom rings are unfavorable due to the bond length requirements. Motif $15_A^Z$ is both geometrically and visually similar to Z15, including the two rings of five atoms and the dual-triangular structure between those rings. (These features of the structure are not highlighted in Figure 8 and are more easily seen in Figure 3 where all the bonds are visible.) Manual alignment of SRO units with CN 16 in the 600 K inherent structure resulted in a motif with topology 1-6-6-3 and VI <0 1 10 5>. This structure is analogous to motif $15_A^Z$ where the two teal atoms in Figure 8 are replaced by a triangle of 3 atoms and is similar to the Frank-Kasper polyhedron Z16 ($D = 0.700$). It is noteworthy that the identification of the structure of this CN 16 motif, coupled with PPM, can resolve the topological discrepancy discussed in Section 3.3 in Ref [37] where CN 16 structures transition from VI <0 1 10 5> to <0 0 12 4> as the glass is further equilibrated; rather than relying on the discontinuous change in topology, PPM comparisons to this motif provide an avenue for quantifying a continuous change in structure during cooling.

*Structure and Chemistry of Select Motifs in $Zr_{50}Cu_{45}Al_5$*

Some motifs in $Zr_{50}Cu_{45}Al_5$ have unique chemical order or correlate with phenomena such as the glass transition; we discuss those motifs in this section.

SRO units with icosahedral and quasi-icosahedral VI are widely reported to play an important role in the structure and dynamics of Zr-Cu-Al MGs [15,56,57]. Motif $12_A^Z$ is an icosahedral motif with VI <0 0 12 0> and extraordinarily high similarity to the geometrically perfect icosahedron. SRO units most like this motif tend to have an abnormally high number of Al atoms in their shell as well as at the center (Figure 7), which is consistent with previous work identifying networks of interpenetrating icosahedra in Zr-Cu-based MGs [27,58–60]. The fraction of atoms assigned to this motif as a function of temperature increases dramatically as the glass goes through its glass transition (Figure 4). In addition, motif $12_A^Z$'s relative energy is dramatically lower than all other motifs, indicating that this motif is the preferred CN 12 structure. These results are in line with other work [3,27,32,61,62] illustrating the unique properties of icosahedra in Zr-Cu-based MGs and confirm that the motif extraction method identifies structures that correlate with properties.

Motif $12_B$ is the other motif with VI <0 0 12 0>, but it is significantly less geometrically icosahedral than motif $12_A^Z$. Many SRO units with VI <0 2 8 2> are most similar to this motif, whereas almost all SRO units that are most similar to motif $12_A^Z$ have VI <0 0 12 0>. We therefore call motif $12_B$ *quasi*-icosahedral. SRO units assigned to motif $12_B$ tend to be Cu- or Al-centered and their shells are Al-poor relative to the base composition. The fraction of SRO units assigned to this motif increases as the material undergoes the glass transition, but not as dramatically as motif $12_A^Z$; however, many more SRO units are most like this motif than motif $12_A^Z$, possibly because the structural constraints in the glass prevents these SRO units from being more perfectly icosahedral. Finally, motif $12_B$ has the 2$^{nd}$ lowest energy of motifs with CN 12, again highlighting the stability of (quasi-)icosahedral structures in $Zr_{50}Cu_{45}Al_5$ MGs.

Motif $12_E$ is a CN 12 motif with VI <0 2 8 2>. SRO units most similar to it tend to be Zr-centered and their shells tend to be Al-rich. The fraction of SRO units assigned to this motif *decreases* through the glass transition, despite the fact that it is more stable than the other non-icosahedral motifs with CN 12. This motif represents many of the Zr-centered, 12-coordinated SRO units, and despite the VI <0 2 8 2> sometimes being considered quasi-icosahedral [26,41,43], SRO units with this structure have the opposite trend with temperature as quasi-icosahedral motifs. We therefore do not consider this motif to be quasi-icosahedral. In addition, motifs $12_E$ and $12_B$ demonstrate that the topological descriptor of VI <0 2 8 2> can mix together SRO units with distinct structures and properties.

Motif $10_A^Z$ is the dominant motif with CN 10 and is similar to Z10. It has VI <0 2 8 0> and has bicapped square antiprism geometry. SRO units with CN 10 are nearly all Cu-centered, as is this motif, and it shows no significant chemical ordering in the shell. Most importantly, the fraction of SRO units assigned to this motif increases significantly with temperature, significantly more strongly than the quasi-icosahedral motif $12_B$, but not nearly as strongly as motif $12_A^Z$. SRO units with VI <0 2 8 0> have been identified as important in various other studies [3,63–67]. The significant increase in relative population of this motif indicates that it may play an important role in the structure of Zr-Cu-based MGs as well, so this motif may deserve more attention in future studies.

Motif $11_A^Z$ is a CN 11 motif with VI <0 2 8 1>. The fraction of SRO units assigned to this motif increases with temperature and it is the only low-energy CN 11 motif. The other two motifs with CN 11 also have VI <0 2 8 1>, demonstrating again that the topological descriptors of VIs can be unable to differentiate between distinct geometric structures.

As a whole, the motifs discussed here provide an abstraction of the structure of $Zr_{50}Cu_{45}Al_5$ MG. Because the abstraction is based on geometry, it may be more understandable than the abstractions of SRO provided by analysis techniques such as Voronoi analysis or common neighbor analysis. Amongst the motifs, we find SRO structures that are both stable (Figure 5) and whose populations increase during cooling (Figure 4). Scoring the SRO units in other glasses with different composition against these motifs using PPM would uncover how SRO changes with *e.g.* composition or introduction of new elements.

*Motif Extraction Method*

We now discuss the advantages and disadvantages of the motif extraction when compared to other techniques that identify structure in MGs. Approaches to identify prototypical features of MG structure fall into two categories: structures can be derived from hypothesized properties of atoms and their bonds, or structure analysis approaches can identify structural features from simulated atomic models. Examples of the former approach include the Frank-Kasper polyhedra and Z-clusters as well as structures generated by the ECP model. These structures have in common various ways of defining and enforcing efficient packing and maximizing atomic number density. However, these approaches often lack chemical information beyond that of the shape or atoms (*e.g.* spherical) and their bond lengths. In addition, the structures that are created are based on known information about MG structure and therefore require significant understanding (or assumptions) of the material structure *a priori*.

On the other hand, in the latter approach, the structural information contained in simulated models is difficult to interpret due to disorder. In the past, motifs have been identified from simulated models by hand by looking at hundreds of SRO units [3,40,68,69], potentially aided by topological characterization techniques such as Voronoi analysis. Unfortunately, this approach is time consuming and is limited by human intuition, which makes it difficult to ensure that all the relevant structures were identified.

Data-driven approaches such as motif extraction offer important alternatives because they remove elements of human limitations. The *cluster alignment* method [18] is similar to our motif extraction technique in that it is data-driven and removes the disorder from the SRO units in simulated models. In the cluster alignment method, a collective alignment first aligns all SRO units with the same CN with respect to one another simultaneously. Then, the pairwise similarity scores are calculated between all pairs of individual SRO units. The collective alignment results in one compromise structure, which highlights the mean structure of the SRO units but masks the structure variability within a CN, especially for motifs that are representative of a small fraction of the SRO units. In addition, collective, all-at-once alignment means that the similarity scores are compromises with alignment to the collective, rather than one-to-one structural distances between individual pairs of structures. As a result, we expect that density based clustering or other machine learning techniques applied to the dissimilarity matrix generated from *cluster alignment* would be less successful than what we report here. Our *motif extraction* method emphasizes structural diversity by calculating accurate pairwise similarity scores using PPM to align individual pairs of SRO units followed by using machine learning clustering to cluster the SRO units into multiple groups with unique structure per CN.

There are drawbacks to motif extraction that may make it unsuitable for certain applications. Motif extraction makes the implicit assumptions that there are a finite number of characteristic motifs that represent the structure and that the model being analyzed contains many SRO units that represent each motif, plus some disorder. The assumption of a small number of motifs will be violated if the disorder in the system is too large, so motif extraction will not be useful for models of a gas and may be not useful for models of colloids at low packing fraction. The assumption of many copies of each motif plus disorder could be violated if the number of atoms in the model is small. Motifs represented by only a few SRO units in a small model may be identified as noise in the HBDSCAN step and therefore not represented by a motif. Neither of these difficulties is present here or in other metallic glass models we have examined in various systems including Al-Sm and Pd-Si, in various model sizes ranging from a few thousand to tens of thousands of atoms, and in various model system methods including molecular dynamics and hybrid reverse Monte Carlo modeling [40].

An additional drawback of motif extraction—in contrast to categorization methods such as VI analysis—is that the process of assigning SRO units to motifs is potentially non-unique and thus somewhat arbitrary. Each SRO unit has a dissimilarity score with respect to each motif, and the vector of these dissimilarity scores can be interpreted as a probability of the structure of the SRO unit being "equal to" the structure of each motif. This means that assigning a SRO unit to a single motif oversimplifies the abstraction of the SRO unit's structure. Put another way, if we make a histogram of all the dissimilarity scores of all SRO units with a given CN aligned to one motif, there is no feature that inspires an obvious cutoff, $D_0$, to assign all SRO units with a dissimilarity less than $D_0$ to the motif. Here, we assigned each SRO unit to the motif to which it is most similar, which provided useful insights.

**Conclusions**

We present a new structure analysis technique called *motif extraction* that leverages point pattern matching, a quantitative structure similarity metric, and machine learning clustering to learn 30 short-range order motifs that describe the structure of a $Zr_{50}Cu_{45}Al_5$ metallic glass quenched via molecular dynamics. These motifs form the basis for the SRO structure in this MG and were learned directly from the atom position data in the simulated model, without human intervention. Of the 30 motifs identified, some motifs are icosahedral or quasi-icosahedral while others are structurally similar to hypothesized close-packed SRO structures (Z-clusters). The motifs that are structurally similar to the Z-clusters form a clear hierarchy of structural order as a function of coordination number, making the placement of an additional atom as the CN number is increased predictable. The icosahedral motif has strong Al chemical order and correlates strongly with the glass transition, and we identified a new motif in this system with CN 10, VI <0 2 8 0>, and bicapped square antiprism geometry that also correlates strongly with the glass transition. We also demonstrate that Voronoi index analysis does not distinguish between SRO units with quantifiably different geometries in some cases, and show that the motif extraction method is a complementary and, in some cases, more robust method to identify important geometric structures in disordered materials.

## Acknowledgements

Development of motif extraction and the application to Zr-Cu-Al metallic glasses by JJM and PMV was supported by the National Science Foundation DMREF program (DMR-1332851 and DMR-1728933). The development of the point pattern matching code by ADB and SP was supported by Air Force Office of Scientific Research (FA9550-17-1-0145). The computing for this research was performed using the compute resources and assistance of the UW-Madison Center for High Throughput Computing (CHTC) in the Department of Computer Sciences. The CHTC is supported by UW-Madison, the Advanced Computing Initiative, the Wisconsin Alumni Research Foundation, the Wisconsin Institutes for Discovery, and the National Science Foundation, and is an active member of the Open Science Grid, which is supported by the National Science Foundation and the U.S. Department of Energy's Office of Science.

## Author Contributions

P.V. and J.M. conceived the idea of motif extraction. J.M. developed the code for motif extraction and carried out the calculations. A.B. and S.P. improved the PPM technique by developing high-throughput capabilities for motif extraction. J.M. wrote the manuscript and all authors contributed to improving it.

## Data Availability

The datasets generated during and/or analyzed during the current study are available from the corresponding author on reasonable request.

## Conflict of Interest

The authors declare that there is no conflict of interest.

Glass Forming Ability of Pd40Ni40P20, Phys. Rev. Lett. 108 (2012) 175501. doi:10.1103/PhysRevLett.108.175501.

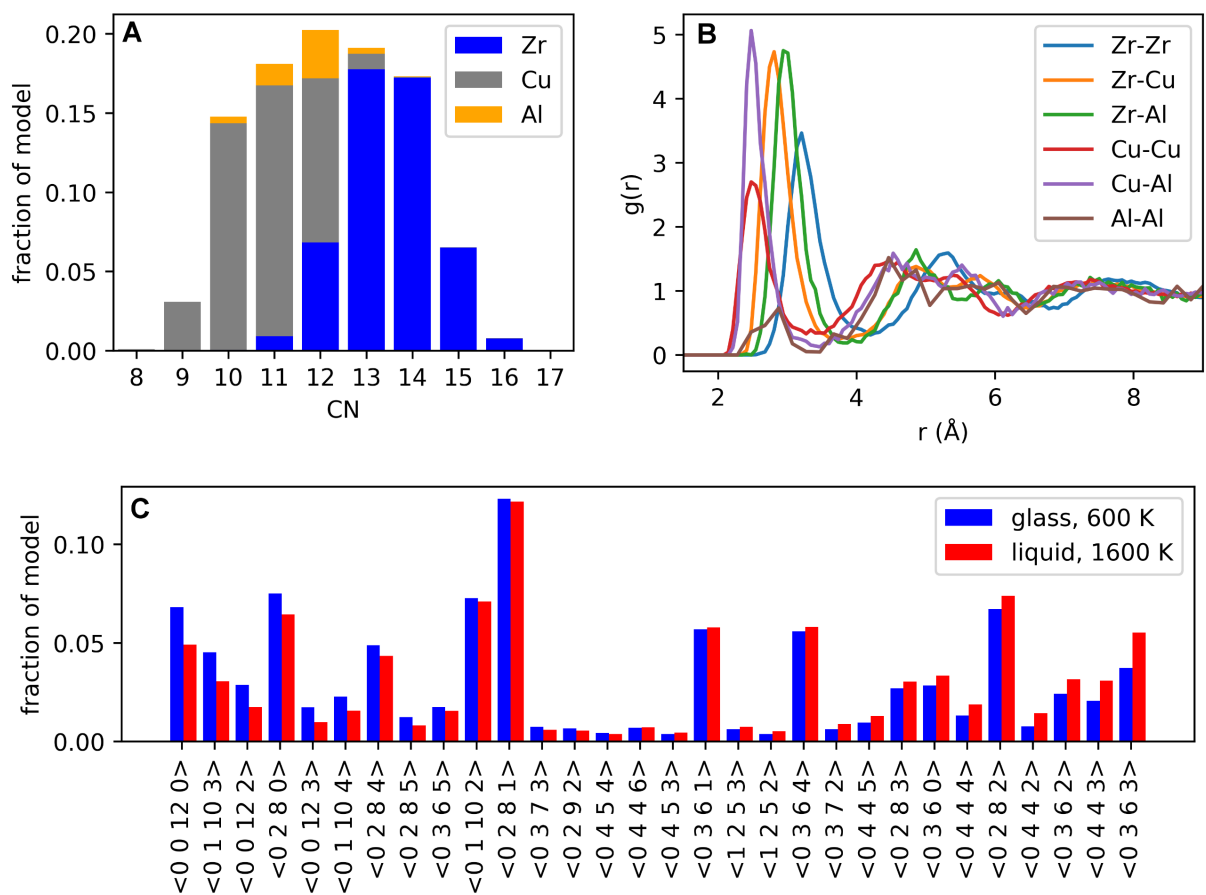

Figure 1: (a) The distribution of coordination numbers in a $Zr_{50}Cu_{45}Al_5$ MG model with 9,826 atoms quenched to 600 K. The colors represent the chemical specie at the center of the SRO units. (b) Partial $g(r)$'s for the same model. (c) The fraction of SRO units in the 600 K and 1600 K models with a given VI. The VIs are sorted on the x-axis by the change in fraction of the model from 1600 K to 600 K.

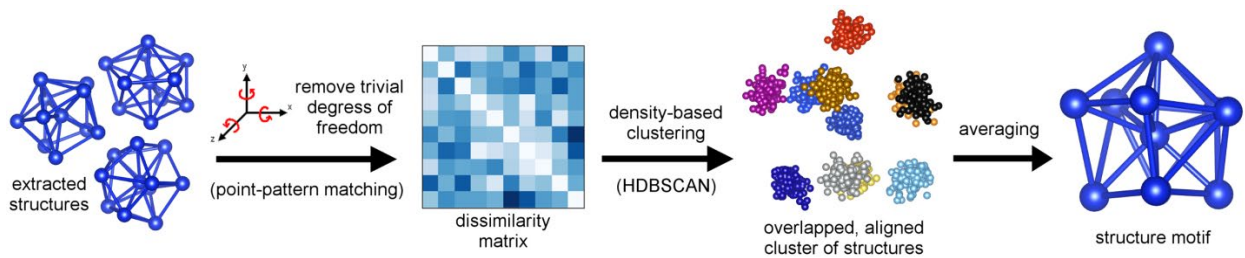

Figure 2: An illustration of the motif extraction method. First, SRO units are extracted from the simulated model. Point-pattern matching aligns all pairs of SRO units and the dissimilarity score, $D$, is calculated. All values of $D$ are combined into a dissimilarity matrix for HDBSCAN, which identifies clusters of similar SRO units. The cluster of SRO units corresponding to motif $10_A^Z$ (see Table 1) is shown. The "bunches" of atoms around the atomic sites are averaged to create the motif. One motif is created for every cluster of similar SRO units identified by HDBSCAN.

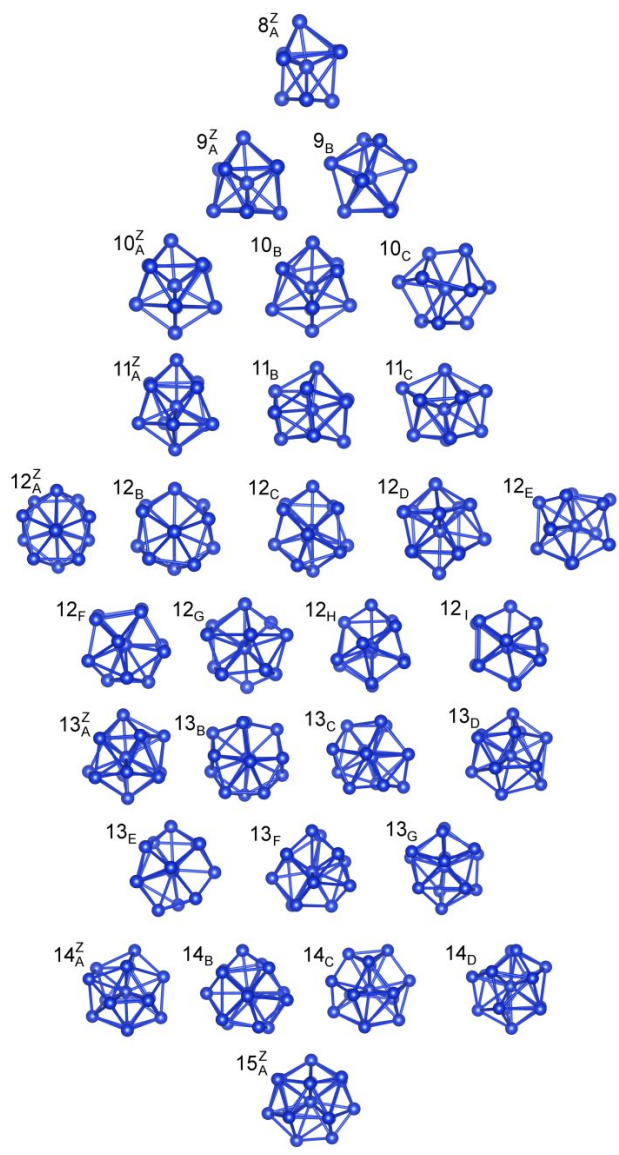

Figure 3: The 30 motifs identified in a $Zr_{50}Cu_{45}Al_5$ MG arranged by CN. Orientations were chosen to illustrate various symmetry elements, if any exist. Note that it is often difficult to show 2D projections that are representative of the 3D structure. Atomic coordinates for these clusters may be found in the SI.

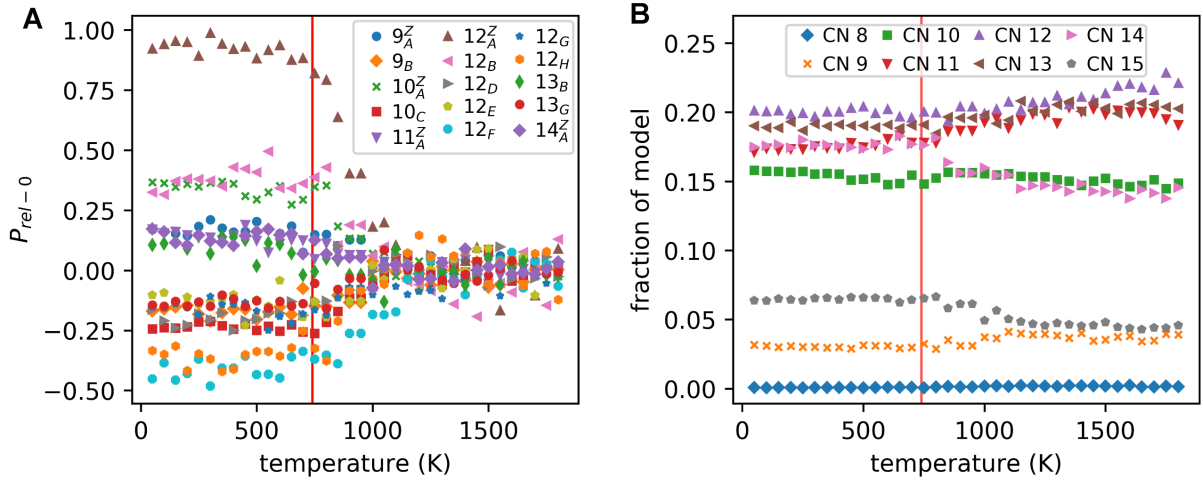

Figure 4: (a) shows a selection of motifs whose population change noticeably during cooling. The y-axis shows the fraction of SRO units for a given CN that were assigned to a motif after normalization as described in the text. Qualitatively, higher values indicate that more SRO units were assigned to a motif than would be expected if the SRO units were distributed evenly among the motifs with the same CN. (b) shows the change in the fraction of each CN with temperature. In both plots, $T_g$ is marked by a solid vertical line.

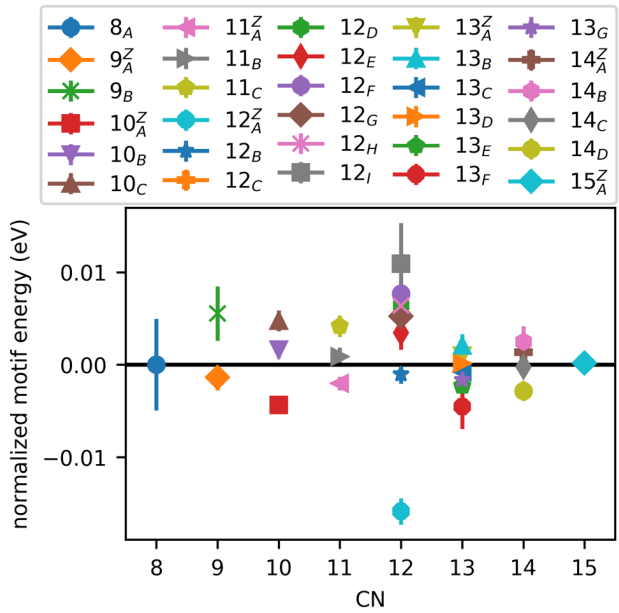

Figure 5: The average energy of SRO units in the 600 K model assigned to each motif after subtracting the average energy of all SRO units with the same CN as the motif. The subtraction allows for direct comparison of the motif energies across CNs.

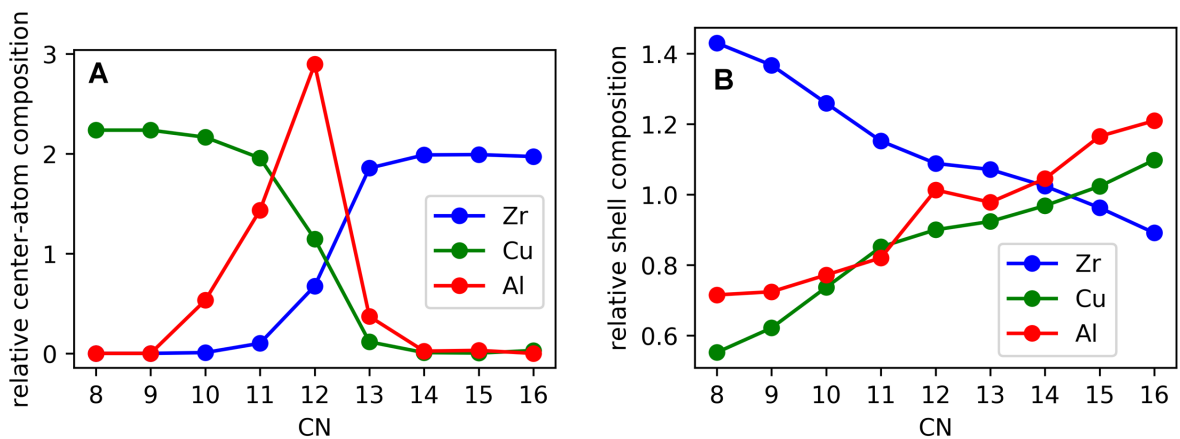

Figure 6: (a) the average center-atom composition of all SRO units in the 600 K model with a given CN, normalized by the composition of the model. (b) the average composition of the atoms in the shell of all SRO units in the 600 K with a given CN, normalized by the composition of the model. Al atoms have an abnormally high tendency to be both at the center and in the shell of SRO units with CN 12.

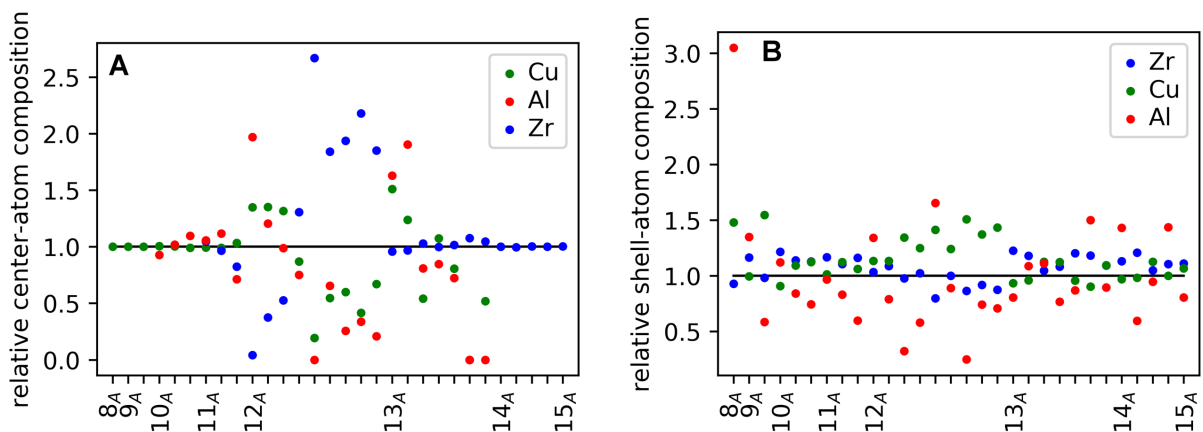

Figure 7: (a) the fraction of Cu-, Al-, and Zr-centered SRO units in the 600 K model assigned to a given motif, normalized to the average composition of all SRO units with the same CN as the motif. Nearly all SRO units with CN 10 are Cu-centered, so the relative center-atom composition of all motifs with CN 10 is 1.0; on the other hand, SRO units with CN 12 can be Cu-, Al-, or Zr-centered so motifs with CN 12 have compositions that differ from the average. (b) the fraction of Cu-, Al-, and Zr- atoms in the shell of each SRO unit assigned to each motif in the 600 K, normalized to the average composition of all SRO units the same CN as the motif. The x-axis ticks are the motifs from Figure 3 in alphanumeric order; only the labels for motifs with subscript A are shown for clarity.

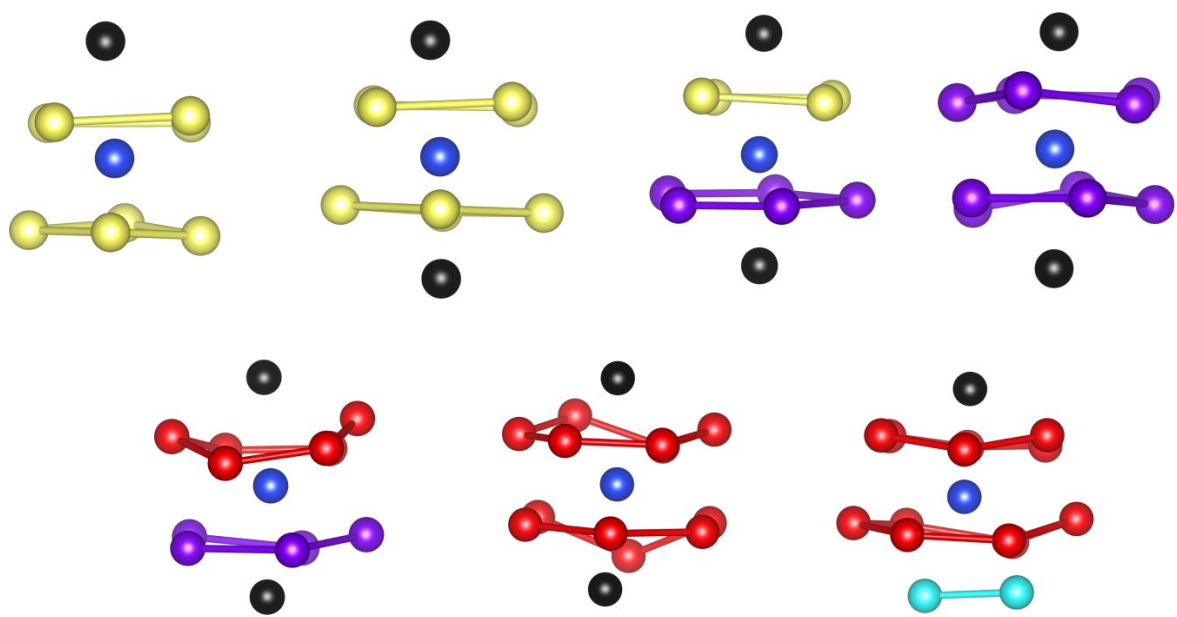

Figure 8: The motif most like the corresponding Z-cluster for each CN, colored to illustrate the planar, ring-like nature of the structures. As the coordination number increases, there is a clear hierarchy of structure and the placement of an additional atom is often predictable. The motifs are colored and bonds are drawn for viewing purposes.

Table 1: The CN, VI, and dissimilarity score (*D*) to the Z-cluster with the same CN for all motifs. See Figure 9 for context for the dissimilarity scores.

| Motif Label | CN | VI | Dissimilarity (*D*) to Z-cluster |
|---|---|---|---|
| $8_A$ | 8 | <0 4 4 0> | 0.823 |
| $9_A{}^Z$ | 9 | <0 3 6 0> | 0.586 |
| $9_B$ | 9 | <0 4 4 1> | 0.961 |
| $10_A{}^Z$ | 10 | <0 2 8 0> | 0.392 |
| $10_B$ | 10 | <0 2 8 0> | 0.538 |
| $10_C$ | 10 | <0 4 4 2> | 1.160 |
| $11_A{}^Z$ | 11 | <0 2 8 1> | 0.381 |
| $11_B$ | 11 | <0 2 8 1> | 0.699 |
| $11_C$ | 11 | <0 2 8 1> | 0.754 |
| $12_A{}^Z$ | 12 | <0 0 12 0> | 0.328 |
| $12_B$ | 12 | <0 0 12 0> | 0.719 |
| $12_C$ | 12 | <0 2 8 2> | 0.933 |
| $12_D$ | 12 | <0 2 8 2> | 0.963 |
| $12_E$ | 12 | <0 2 8 2> | 1.157 |
| $12_F$ | 12 | <0 2 8 2> | 1.157 |
| $12_G$ | 12 | <0 2 8 2> | 1.165 |
| $12_H$ | 12 | <0 3 6 3> | 1.250 |
| $12_I$ | 12 | <0 4 4 4> | 1.176 |
| $13_A{}^Z$ | 13 | <0 1 10 2> | 0.418 |
| $13_B$ | 13 | <0 1 10 2> | 0.555 |
| $13_C$ | 13 | <0 1 10 2> | 0.679 |
| $13_D$ | 13 | <0 1 10 2> | 0.598 |
| $13_E$ | 13 | <0 2 8 3> | 1.046 |
| $13_F$ | 13 | <0 3 6 4> | 1.128 |
| $13_G$ | 13 | <0 3 6 4> | 1.010 |
| $14_A{}^Z$ | 14 | <0 2 8 4> | 0.683 |
| $14_B$ | 14 | <0 2 8 4> | 1.017 |
| $14_C$ | 14 | <0 0 12 4> | 0.756 |
| $14_D$ | 14 | <0 1 10 3> | 0.806 |
| $15_E{}^Z$ | 15 | <0 1 10 4> | 0.475 |
| Z8 | 8 | <0 4 4 0> | 0. |
| Z9 | 9 | <0 3 6 0> | 0. |
| Z10 | 10 | <0 2 8 0> | 0. |
| Z11 | 11 | <0 2 8 1> | 0. |
| Z12 | 12 | <0 0 12 0> | 0. |
| Z13 | 13 | <0 1 10 2> | 0. |
| Z14 | 14 | <0 0 12 2> | 0. |
| Z15 | 15 | <0 0 12 3> | 0. |

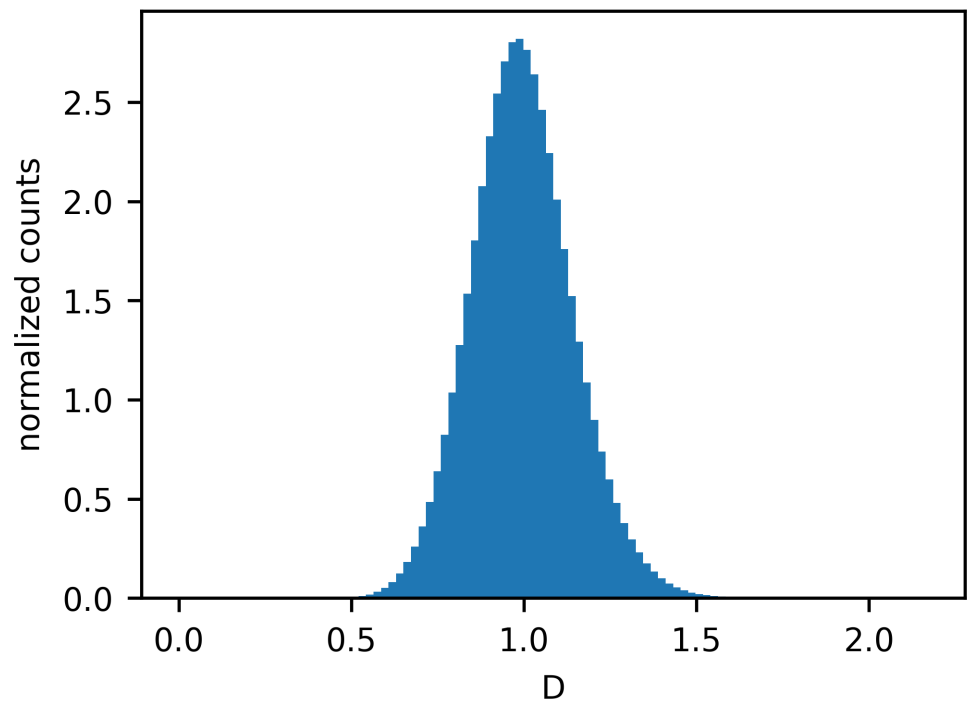

Figure 9: A histogram of all dissimilarity scores, *D*, in the 600 K model calculated by motif extraction.